\documentclass[twocolumn,showpacs,preprintnumbers,amsmath,amssymb,prl]{revtex4-1}

\usepackage{graphicx}
\usepackage{dcolumn}
\usepackage{bm}
\usepackage{tabularx}
\usepackage{amsmath}

\begin{document}

\title{Absence of magnetic thermal conductivity in the quantum spin liquid candidate YbMgGaO$_4$}

\author{Y. Xu,$^1$ J. Zhang,$^1$ Y. S. Li,$^{2,3}$ Y. J. Yu,$^1$ X. C. Hong,$^1$ Q. M. Zhang,$^{2,4}$ and S. Y. Li$^{1,4,*}$}

\affiliation{$^1$State Key Laboratory of Surface Physics, Department of Physics,
and Laboratory of Advanced Materials, Fudan University, Shanghai 200433, China\\
$^2$Department of Physics, Renmin University of China, Beijing 100872, China\\
$^3$Experimental Physics VI, Center for Electronic Correlations and Magnetism,
University of Augsburg, 86159 Augsburg, Germany\\
$^4$Collaborative Innovation Center of Advanced Microstructures, Nanjing 210093, China
}

\date{\today}

\begin{abstract}
We present the ultra-low-temperature specific heat and thermal conductivity measurements on the single crystals of YbMgGaO$_4$, which was recently argued to be a promising candidate for quantum spin liquid (QSL). In the zero magnetic field, a large magnetic contribution of specific heat is observed, and exhibits a power-law temperature dependence ($C_m \sim T^{0.74}$). On the contrary, we do not observed any significant contribution of thermal conductivity from magnetic excitations. In magnetic fields $H \ge$ 6 T, the exponential $T$-dependence of $C_m$ and the enhanced thermal conductivity indicate a magnon gap of the fully-polarized state. The absence of magnetic thermal conductivity at the zero field in this QSL candidate puts a strong constraint on the theories of its ground state.
\end{abstract}


\maketitle

The notion of the quantum spin liquid (QSL) reentered the view of researchers in 1987 \cite{Anderson1}, fourteen years after it was first proposed by Anderson when he tackled the possibility of a peculiar destruction of magnetism exhibited by spins in a triangular lattice \cite{Anderson2}. Ever since then, the passion for searching candidate materials that may harbor such an exotic state of matter has never cooled down \cite{search1,k-salt,k-salt specific heat,k-salt kappa,search2,herbertsmithite0,herbertsmithite1,herbertsmithite2,herbertsmithite3,ZnCu neutron,ZnCu NMR,dmit,dmit kappa,dmit specific heat,ZnCu SO4}. In a QSL, a macroscopic number of spins are entangled but can evade symmetry-breaking long-range magnetic order with the help of geometrical frustration, and remain fluid-like even in the zero-temperature limit. Instead of adopting a static arrangement, the spins fluctuate perpetually \cite{PALee,Balents}.

As the QSL state was firmly established in one-dimensional spin systems \cite{1D1,1D2}, realizing QSLs in two- and three-dimensional systems has been pursued extensively. Of specific interest has been the spin-1/2 triangular- and kagome-lattice Heisenberg antiferromagnets, in which the former one is the very prototype of a QSL in Anderson's resonating-valence-bond model \cite{Anderson1,Anderson2,Baskaran}. After the wave of research on the star systems like $\kappa$-(BEDT-TTF)$_2$Cu$_2$(CN)$_3$ \cite{k-salt,k-salt specific heat,k-salt kappa}, ZnCu$_3$(OH)$_6$Cl$_2$ \cite{herbertsmithite0,herbertsmithite1,herbertsmithite2,herbertsmithite3,ZnCu neutron,ZnCu NMR}, and EtMe$_3$Sb[Pd(dmit)$_2$]$_2$ \cite{dmit,dmit kappa,dmit specific heat}, the newly discovered YbMgGaO$_4$ was argued to be a promising candidate for QSL \cite{srep,PRL,muSR,neutron SY,neutron2}. No indication of magnetic ordering was observed in specific heat measurements on polycrystalline YbMgGaO$_4$ down to 60 mK, far below the Curie-Weiss temperature $\theta_W$ $\approx $ 4 K \cite{srep}. A broad continuum of spin excitations, which is a hallmark of the QSL state, was observed in neutron scattering measurements, confirming YbMgGaO$_4$ to be a highly promising QSL candidate \cite{neutron SY,neutron2}. Furthermore, the ground state of YbMgGaO$_4$ was proposed to be a gapless U(1) QSL with a spinon Fermi surface, which was evidenced by the temperature dependence of the specific heat ($C_m$ $\sim T^{2/3}$) \cite{srep,neutron2}, the muon spin relaxation ($\mu$SR) results \cite{muSR}, and the crucial features in the inelastic neutron scattering spectrum \cite{neutron SY}. As for the mechanism to stabilize a QSL ground state on the triangular lattice of YbMgGaO$_4$, there are two potential ones: while the electron-spin resonance measurement ascribes the QSL physics to the anisotropy of the nearest-neighbor spin interaction \cite{PRL}, the neutron scattering study identifies the next-nearest-neighbor interactions in the presence of planar anisotropy as key ingredients for the QSL formation \cite{neutron2}.

To understand the nature of a QSL, knowledge of the low-lying elementary excitations would be of primary importance. Ultra-low-temperature specific heat and thermal conductivity measurements have proven to be powerful means in the study of low-lying excitations in QSL candidates \cite{k-salt specific heat,k-salt kappa,dmit specific heat,dmit kappa}. Although the gapless feature of the low-energy excitations was reported by the specific heat measurement in $\kappa$-(BEDT-TTF)$_2$Cu$_2$(CN)$_3$ \cite{k-salt specific heat}, the thermal conductivity result implied a possibility of a tiny gap opening \cite{k-salt kappa}. In the case of EtMe$_3$Sb[Pd(dmit)$_2$]$_2$, both measurements indicated the existence of gapless spin excitations \cite{dmit specific heat,dmit kappa}.

In this Letter, we report the ultra-low-temperature specific heat and thermal conductivity measurements on high-quality YbMgGaO$_4$ single crystals. In the zero magnetic field, a large magnetic contribution with a power-law temperature dependence ($C_m \sim T^{0.74}$) is observed in the specific heat. However, no significant contribution from magnetic excitations is detected in the thermal conductivity. In magnetic fields $H \ge$ 6 T, the behaviors of the specific heat and the thermal conductivity are consistent with a fully-polarized state. We discuss the origin of the absence of magnetic thermal conductivity in this QSL candidate.

The high-quality single crystals of YbMgGaO$_4$ used in this work, as well as the non-magnetic isostructural material LuMgGaO$_4$, were grown by the floating zone technique \cite{PRL}. The specific heat of the YbMgGaO$_4$ single crystal was measured from 0.05 to 3 K in a physical property measurement system (PPMS, Quantum Design) equipped with a small dilution refrigerator. The YbMgGaO$_4$ sample for the thermal conductivity measurements was cut to a rectangular shape of dimensions 2.50 $\times$ 0.81 mm$^2$ in the $ab$ plane, with a thickness of 0.30 mm along the $c$ axis. Contacts were made directly on the sample surfaces with silver epoxy. An annealing process was conducted at 400 $^{\circ}$C for 30 minutes to gain a better contact. The thermal conductivity was measured in a dilution refrigerator, using a standard four-wire steady-state method with two RuO$_2$ chip thermometers, calibrated $in$ $situ$ against a reference RuO$_2$ thermometer. Magnetic fields were applied along the $c$ axis for both the specific heat and thermal conductivity measurements. For comparison, the thermal conductivity of the LuMgGaO$_4$ single crystal was also measured on a sample with dimensions of 1.48 $\times$ 0.78 $\times$ 0.31 mm$^3$.

\begin{figure}
\includegraphics[clip,width=8.7cm]{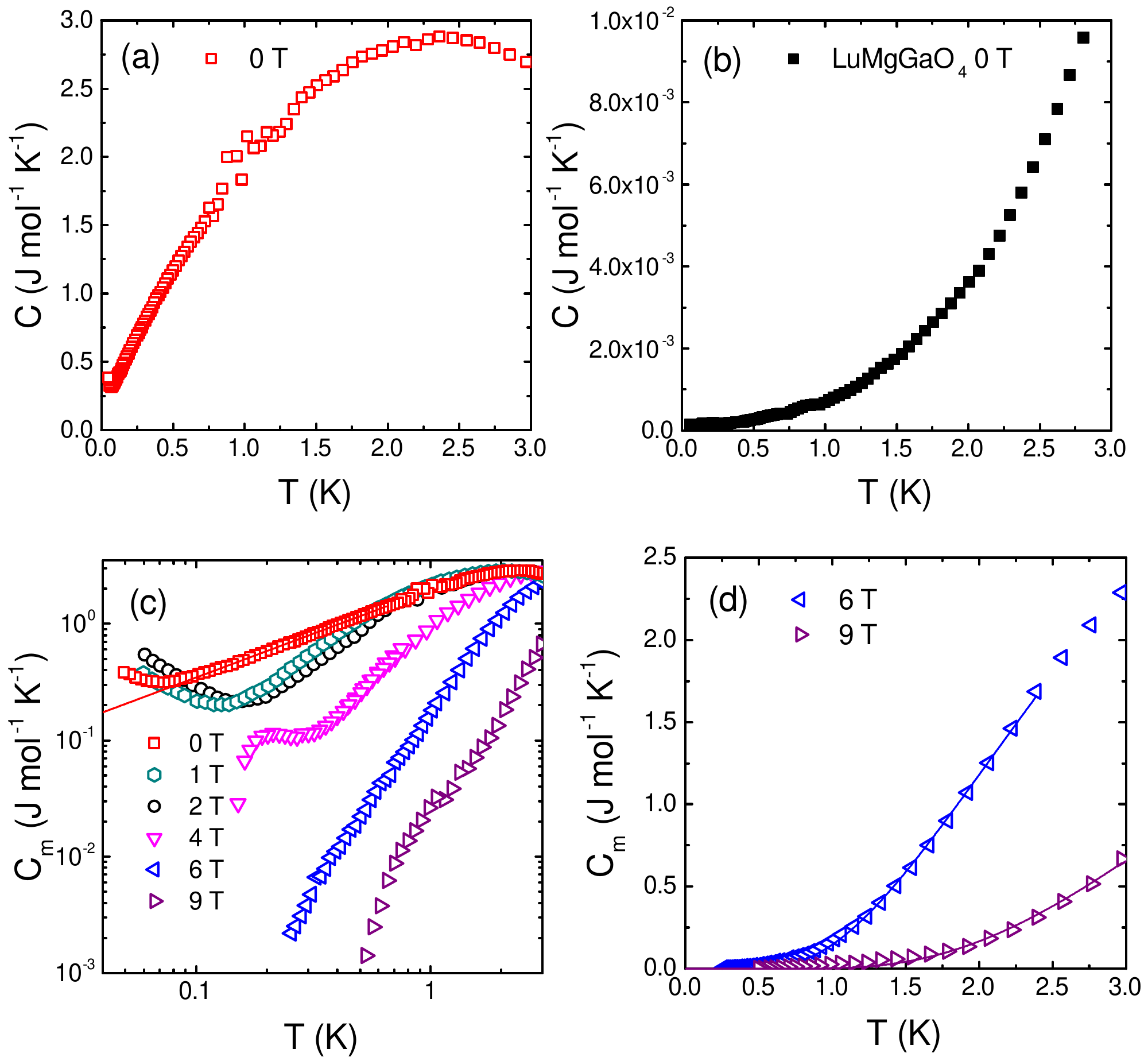}
\caption{ (a) The specific heat of the YbMgGaO$_4$ single crystal at the zero magnetic field $H$ = 0 T. (b) The specific heat of the LuMgGaO$_4$ polycrystalline sample at $H$ = 0 T (data from Ref. \cite{srep}). (c) The magnetic specific heat of the YbMgGaO$_4$ single crystal at various magnetic fields up to 9 T in a log-log scale. The magnetic specific heat is extracted by subtracting the specific heat of LuMgGaO$_4$ from that of YbMgGaO$_4$. The solid line is the fit of the 0 T data to $C_m$ = $c$$T^{\beta}$ between 0.10 and 0.65 K. (d) The specific heat  of the YbMgGaO$_4$ single crystal at $H$ = 6 and 9 T. The solid lines are the fits to $C_m$ = $d$$e^{-\Delta/k_BT}$, in which $\Delta$ is the magnon gap in the fully-polarized state.
 }
\end{figure}

Figure 1(a) shows the specific heat of the YbMgGaO$_4$ single crystal at the zero magnetic field $H$ = 0 T, which is nearly identical to the data of the polycrystalline sample we measured previously \cite{srep}. Figure 1(b) shows the specific heat of the non-magnetic counterpart LuMgGaO$_4$ polycrystalline sample (data from Ref. \cite{srep}). It can be clearly seen that the magnitude of the specific heat of YbMgGaO$_4$ is far beyond that of LuMgGaO$_4$ in our temperature range (0.05 $\sim$ 3 K). As described in Ref. \cite{srep}, the specific heat of LuMgGaO$_4$ well follows the Debye law with a Debye temperature $\sim$ 151 K. The magnetic specific heat ($C_m$) of YbMgGaO$_4$ can be extracted by subtracting the lattice contribution, i.e., the specific heat of LuMgGaO$_4$, from that of YbMgGaO$_4$. The $C_m$ of YbMgGaO$_4$ in zero and finite magnetic fields up to 9 T are plotted in Fig. 1(c) in a log-log scale. The feature at the lowest temperatures comes from the Schottky contribution. As seen in Fig. 1(c), the zero-field specific heat of YbMgGaO$_4$ can be well fitted by $C_m$ = $c$$T^{\beta}$ with $\beta$ = 0.74 (fitting range 0.10 $\sim$ 0.65 K). This value coincides with the value reported previously \cite{srep,neutron2}, and is close to 2/3. As analyzed in  Ref. \cite{neutron SY}, a gapless QSL with a spinon Fermi surface would give a spinon specific heat $C_m$ $\sim T$, which is further corrected to $C_m$ $\sim T^{2/3}$ if there is strong U(1) gauge fluctuations \cite{neutron SY}.

Although the zero-field specific heat of the YbMgGaO$_4$ single crystal can be fitted into a theoretical framework satisfactorily, the picture is rather complicated under magnetic fields. As seen in Fig. 1(c), the magnetic field rapidly suppresses the $C_m$. Under magnetic fields, the temperature dependence of the $C_m$ gradually turns into an exponential one, as seen in Fig. 1(d). Such an exponential $C_m(T)$ is attributed to the magnons with a gap in a fully-polarized state. This fully-polarized state for YbMgGaO$_4$ with very small exchange couplings is evidenced in the magnetization measurements, from which the magnetization tends to saturate above $H$ = 6 T in $H \parallel c$ and $T \sim$ 2 K \cite{PRL,neutron SY}.  In Fig. 1(d), we fit the 6 T and 9 T data with $C_m$ = $d$$e^{-\Delta/k_BT}$, in which $\Delta$ is the magnon gap in the fully-polarized state. The fittings give $\Delta$ = 4.17 K, $d$ = 9.50 J mol$^{-1}$ K$^{-1}$ (fitting range 0.81 $\sim$ 2.39 K) and $\Delta$ = 8.26 K, $d$ = 10.4 J mol$^{-1}$ K$^{-1}$ (fitting range 1.24 $\sim$ 2.97 K), respectively. With $J_{\pm\pm}$ = 0.2$J_{zz}$, and $J_{z\pm}$ = 0.3$J_{zz}$, and a fitting formula from Ref. \cite{YDLiPRB}, we calculate the magnon gap $\Delta$ to be 0.41 meV (4.69 K) for $H =$ 6 T and 0.73 meV (8.40 K) for $H =$ 9 T along the $c$ axis. These calculated gap values are in good match with the fitted ones from our specific heat data.

\begin{figure}
\includegraphics[clip,width=6.6cm]{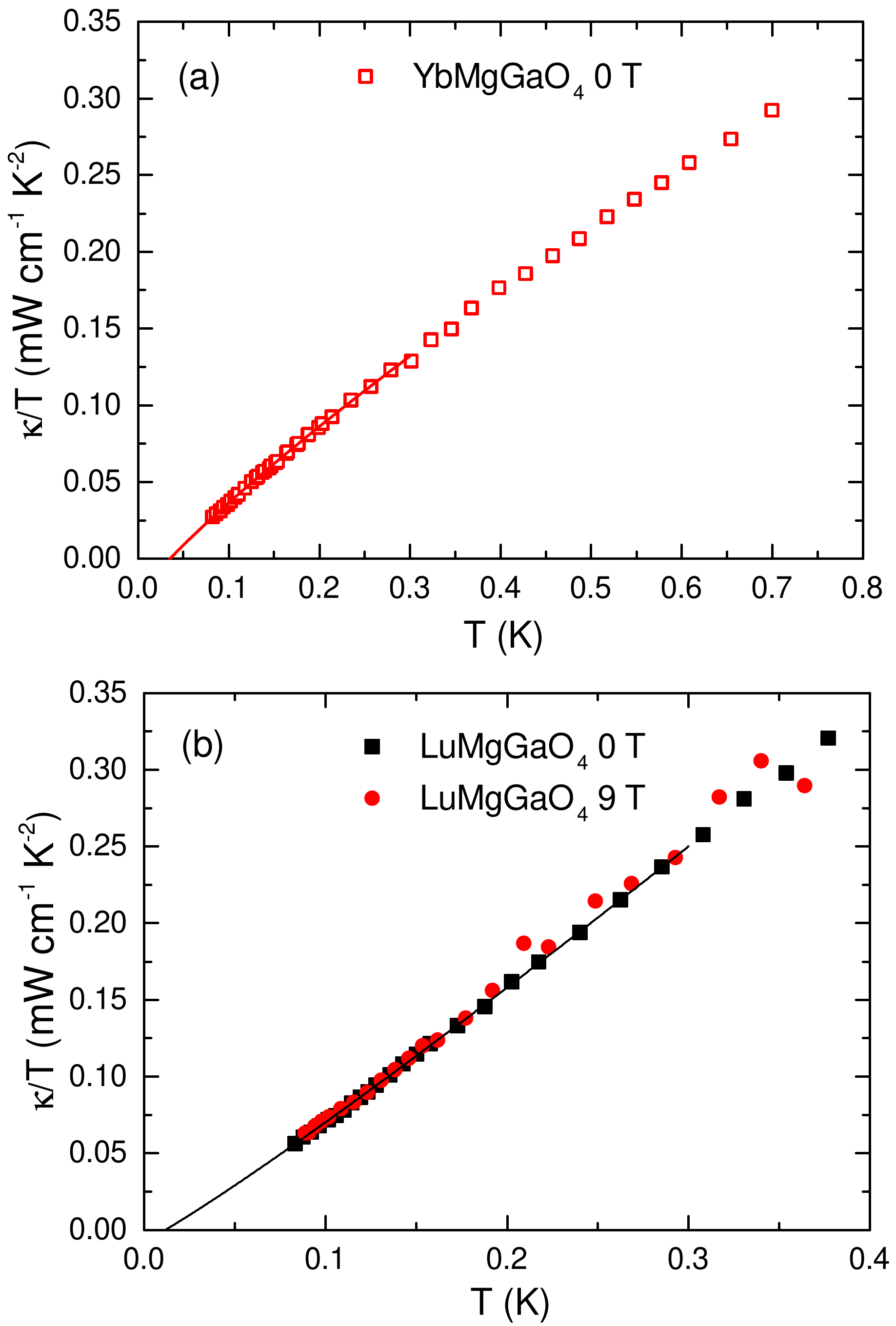}
\caption{The in-plane thermal conductivity of (a) the YbMgGaO$_4$ single crystal at $H$ = 0 T and (b) the LuMgGaO$_4$ single crystal at $H$ = 0 and 9 T. The solid lines are the fits to the data below 0.3 K to $\kappa/T$ = $a$ + $bT^{\alpha-1}$. Note that the applying of a 9 T magnetic field has no effect on the thermal conductivity of non-magnetic LuMgGaO$_4$.}
\end{figure}

\begin{figure}
\includegraphics[clip,width=6.65cm]{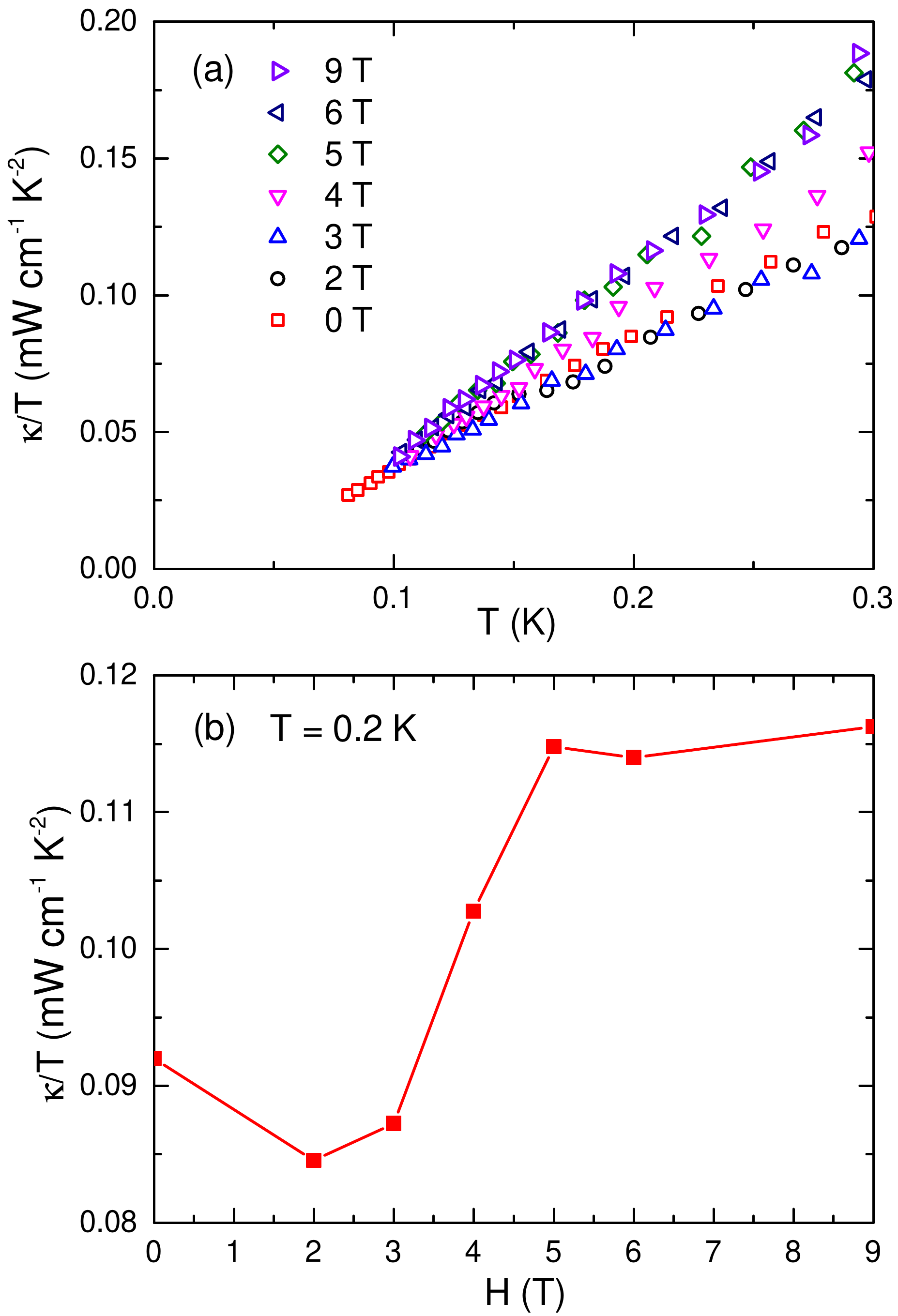}
\caption{(a) The in-plane thermal conductivity of the YbMgGaO$_4$ single crystal at various magnetic fields up to 9 T. (b) Field dependence of the $\kappa/T$ at 0.2 K. The $\kappa/T$ first decreases slightly for $H$ $<$ 2 T, then increases sharply (by about 35\%) for 2 T $<$ $H$ $<$ 5 T, and finally saturates for $H$ $>$ 5 T. The saturated thermal conductivity above 5 T is purely attributed to phonons, without scattering by magnetic excitations.}
\end{figure}

Thermal conductivity measurement is highly advantageous in probing the elementary excitations in QSL candidates, since it is only sensitive to itinerant excitations and is not complicated by the Schottky contribution as observed in the specific heat measurement \cite{kappa calc1,k-salt kappa}. Figure 2(a) shows the in-plane thermal conductivity of the YbMgGaO$_4$ single crystal at $H$ = 0 T. For comparison, the in-plane thermal conductivity of the non-magnetic counterpart LuMgGaO$_4$ single crystal at $H$ = 0 T is also plotted in Fig. 2(b). In a solid, the contributions to thermal conductivity may come from various quasiparticles, such as electrons, phonons, magnons, and spinons. For non-magnetic compounds, the thermal conductivity at very low temperature can usually be fitted to $\kappa$ = $aT$ + $bT^{\alpha}$, in which the two terms $aT$ and $bT^{\alpha}$ represent contributions from electrons and phonons, respectively \cite{specular reflection1,specular reflection2}. Because of the specular reflections of phonons at the sample surfaces, the power $\alpha$ in the second term is typically between 2 and 3 \cite{specular reflection1,specular reflection2}. The fitting of the data below 0.3 K for LuMgGaO$_4$ gives $\kappa_0/T \equiv a =$ -0.007 $\pm$ 0.002 mW K$^{-2}$ cm$^{-1}$, and $\alpha$ = 2.09 $\pm$ 0.02. Comparing with our experimental error bar $\pm$ 0.005 mW K$^{-2}$ cm$^{-1}$, the $\kappa_0/T$ of LuMgGaO$_4$ at zero field is virtually zero. This is reasonable, since LuMgGaO$_4$ is an insulator. For YbMgGaO$_4$ with a triangular lattice of spins, one may expect a significant contribution to thermal conductivity by magnetic excitations due to its large $C_m$. However, with the first glance at the raw data in Fig. 2(a), the magnitude of its $\kappa$ is only half of that of LuMgGaO$_4$ in Fig. 2(b), although the two samples have comparable cross-section area (thus the mean free path of the phonons in the boundary scattering limit). We also fit the zero-field data of YbMgGaO$_4$ below 0.3 K to $\kappa/T$ = $a$ + $bT^{\alpha-1}$, which gives $\kappa_0/T \equiv a$ = -0.025 $\pm$ 0.002 mW K$^{-2}$ cm$^{-1}$ and $\alpha$ = 1.85 $\pm$ 0.02.

A negative $\kappa_0/T$ has no physical meaning, and the power $\alpha$ is abnormally lower than 2 for YbMgGaO$_4$. Since the specific heat measurements indicate the existence of a sufficient amount of magnetic excitations in YbMgGaO$_4$, it is reasonable to assume that the phonons are scattered not only by the sample boundaries, but also by these magnetic excitations. To verify this assumption, we examine the thermal conductivity of YbMgGaO$_4$ in magnetic fields up to 9 T, as plotted in Fig. 3(a). While the applying of a 9 T field has no effect on the $\kappa$ of non-magnetic LuMgGaO$_4$, as seen in Fig. 2(b), the field has a significant effect on the $\kappa$ of YbMgGaO$_4$. Figure 3(b) plots the field dependence of the $\kappa/T$ at 0.2 K. The $\kappa/T$ first decreases slightly for $H$ $<$ 2 T, then there is a sharp increase (by about 35\%) between 2 and 5 T, and it finally saturates for $H >$ 5 T. While the magnetic state of YbMgGaO$_4$ in the intermediate fields (0 $< H <$ 5 T) is rather complex, it simply tends to become a fully-polarized state for $H$ $>$ 5 T at such low temperatures, as evidenced by previous magnetization \cite{PRL,neutron SY} and our current specific heat measurements. In the fully-polarized state with the magnon gap of several Kelvins, there are almost no magnetic excitations to scatter phonons below 0.3 K, therefore the $\kappa$ of YbMgGaO$_4$ at $H >$ 5 T is purely contributed by phonons. Indeed, both the magnitude and the temperature dependence of $\kappa$ for YbMgGaO$_4$ at $H >$ 5 T are more closer to those of LuMgGaO$_4$. At lower fields, it is the additional scattering of phonons by magnetic excitations that suppresses the $\kappa$ and gives the abnormal temperature dependence of $\kappa$ and the un-physical negative $\kappa_0/T$ for YbMgGaO$_4$.

The large $C_m$ and its temperature dependence ($C_m$ $\sim T^{2/3}$) suggest a gapless U(1) QSL with a spinon Fermi surface \cite {neutron SY}. Neutron scattering measurement also observed diffusive spin excitations above 0.1 meV, indicating the particle-hole excitation of a spinon Fermi surface \cite{neutron SY}. In this context, it is quite surprising that we do not observe any significant magnetic contribution to the $\kappa$ of YbMgGaO$_4$. By contrast, EtMe$_3$Sb[Pd(dmit)$_2$]$_2$ has a $\kappa_0/T$ as big as 2 mW K$^{-2}$ cm$^{-1}$ \cite{dmit kappa}. This means that either a) the presumed gapless spinons do not exist in YbMgGaO$_4$ and the large $C_m$ has some other magnetic origin; or b) the gapless spinons do exist but for some reason they do not conduct heat significantly in YbMgGaO$_4$.

In the case that YbMgGaO$_4$ does have a QSL state with gapless spinons, there are further two possibilities. One is that the ground state can be described as a gapless U(1) QSL \cite{PRL,muSR,neutron SY}. In this scenario, the low-energy spinons are no longer well-defined Laudau quasiparticles, and the simple kinetic formula is invalid. Note that this may not be true for YbMgGaO$_4$ at $T >$ 0.1 K, since the gauge field scattering might only take effect at lower temperatures due to its very low isotropic Heisenberg coupling $J_0 \sim$ 1.5 K \cite{PRL}. Nevertheless, considering the strong U(1) gauge fluctuations, a theoretical formula is derived for a gapless U(1) QSL with a spinon Fermi surface in the clean limit as below \cite{kappa calc1,kappa calc2}:
\begin{equation}
\frac{\kappa}{T} = \frac{k_{B}^{2}}{\hbar}(\frac{{\epsilon}_{F}}{k_{B}T})^{\frac{2}{3}}\frac{1}{d},
\end{equation}
where ${\epsilon}_{F}$ is the spinon Fermi energy, and $d$ is the interlayer distance. Taking ${\epsilon}_{F}$ $\approx J_0$ $\sim$ 1.5 K \cite{k-salt kappa,PRL} and $d$ = 25 \AA\ \cite{srep}, we estimate $\kappa/T \approx$ 0.044 mW K$^{-2}$ cm$^{-1}$ at $T$ = 0.1 K. This value is even higher than the total $\kappa/T$ we measured, and there is also no visible $T^{-2/3}$-dependent spinon thermal conductivity $\kappa/T$ on top of the normal phonon contribution in Fig. 2(a). It is not clear whether the impurity scattering will further reduce this estimation \cite{kappa calc1,k-salt kappa}, and cause the absence of magnetic thermal conductivity in YbMgGaO$_4$. We also notice another calculation of the thermal conductivity for a spinon Fermi surface coupled to a U(1) gauge field, which gives $\kappa/T \approx A_{xx}T^{-1/2+5\epsilon/4} + A_{yy}T^{1/2+3\epsilon/4}$ in an intermediate temperature regime \cite{controlled c,controlled kappa}. However, if a phonon term $bT^{\alpha-1}$ is added to fit the total $\kappa/T$, there are too many parameters for us to make a quantitative analysis.

Another possibility is that the gapless spinons in the QSL ground state of YbMgGaO$_4$ are still well-defined Laudau quasiparticles. Then we try to find out the origin of the absence of spinon thermal conductivity by estimating their mean free path. According to the kinetic formula, the thermal conductivity is written as $\kappa_m$ = $\frac{1}{3}C_m v_F l$, where $C_m$, $v_F$ and $l$ are the specific heat, Fermi velocity and the mean free path of spinons, respectively. A large $C_m$ and a negligible contribution to thermal conductivity might come from the reduction of $v_F$ or/and $l$. By comparing Fig. 3a (intensity contour plot of spin excitation spectrum along the high-symmetry momentum directions) and Fig. S3b (calculated dynamic spin structure factor along high symmetry points) of Ref. \cite{neutron SY}, we get $v_F$ = 1.82 $\times$ 10$^2$ m/s. Even if the $\kappa/T$ at 0.1 K is totally contributed by spinons, $l$ would only be 8.6 \AA, about 2.5 times of the inter-spin distance ($\sim$ 3.4 \AA). For comparison, the gapless excitations have an $l$ as long as $\sim$ 1000 inter-spin distance in EtMe$_3$Sb[Pd(dmit)$_2$]$_2$ \cite{dmit kappa}.  For EtMe$_3$Sb[Pd(dmit)$_2$]$_2$, a linear term of $\gamma$ = 20 mJ K$^{-2}$ mol$^{-1}$ \cite{dmit specific heat} in the specific heat and a linear term of $\kappa_0/T$ = 2 mW K$^{-2}$ cm$^{-1}$ \cite{dmit kappa} in the thermal conductivity indicate the presence of highly mobile gapless magnetic excitations with an extremely long $l$. In the case that the gapless spinons do exist in YbMgGaO$_4$, although its $C_m$ is one order of magnitude larger than that of EtMe$_3$Sb[Pd(dmit)$_2$]$_2$, the small $v_F$ and the extremely short $l$ might be the reason why these spinons do not contribute significantly to the thermal conductivity at the zero magnetic field. One possible mechanism of the spinon localization may be the disorder of Mg$^{2+}$-Ga$^{3+}$ sites (random occupation) in the double layers of Mg/GaO$_5$ triangular bipyramids \cite{srep}.

For another triangular-lattice QSL candidate $\kappa$-(BEDT-TTF)$_2$Cu$_2$(CN)$_3$, the specific heat measurement gives a linear term of $\gamma$ = 15 mJ K$^{-2}$ mol$^{-1}$ \cite{k-salt specific heat}. However, the specific-heat data are plagued by a very large nuclear Schottky contribution below 1 K, which might lead to ambiguity \cite{k-salt kappa}. The magnetic part of the thermal conductivity of $\kappa$-(BEDT-TTF)$_2$Cu$_2$(CN)$_3$ exhibits an exponential temperature dependence and gives negligible $\kappa_0/T$, which was interpreted as evidence of a gapped QSL \cite{k-salt kappa}. An alternative explanation to reconcile the specific heat and thermal conductivity of $\kappa$-(BEDT-TTF)$_2$Cu$_2$(CN)$_3$ is that the gapless spin excitations may be localized due to the inhomogeneity \cite{inhomogeneity}. Here for YbMgGaO$_4$, its ground state apparently can not be described by the gapped QSL due to the large $C_m$ down to 0.1 K. More low-energy experimental techniques, such as nuclear magnetic resonance, are highly desired to determine whether its ground state is a gapless QSL with localized spinons.

In summary, we have measured the ultra-low-temperature specific heat and thermal conductivity of the YbMgGaO$_4$ single crystals. The large magnetic specific heat $C_m$ down to 0.1 K and its power-law temperature dependence ($C_m \sim T^{0.74}$) suggest gapless magnetic excitations. The exponential $C_m$ at fields above 6 T indicates a fully-polarized state. The thermal conductivity reveals no significant positive contribution from magnetic excitations. Instead, it is dominated by phonons, and the additional scattering of phonons by magnetic excitations at low fields reduces its value. The absence of magnetic thermal conductivity at the zero field in YbMgGaO$_4$ puts a strong constraint on the theories of its ground state.

We thank Shiliang Li, Hong Yao, Weiqiang Yu, and Yi Zhou for helpful discussions. This work is supported by the Ministry of Science and Technology of China (Grant No: 2015CB921401, 2016YFA0300503, and 2016YFA0300504), the Natural Science Foundation of China, the NSAF (Grant No: U1630248), the Program for Professor of Special Appointment (Eastern Scholar) at Shanghai Institutions of Higher Learning, and STCSM of China (No. 15XD1500200). \\

\noindent $^*$ E-mail: shiyan$\_$li@fudan.edu.cn.

\end{document}